
\input harvmac
\input amssym.def
\input amssym
\baselineskip 14pt
\magnification\magstep1

\parskip 6pt
\newdimen\itemindent \itemindent=32pt
\def\textindent#1{\parindent=\itemindent\let\par=\resetpar%
\indent\llap{#1\enspace}\ignorespaces}

\let\oldpar=\par
\def\resetpar{\oldpar\parindent=20pt\let\par=\oldpar}

\font\ninerm=cmr9 \font\ninesy=cmsy9
\font\eightrm=cmr8 \font\sixrm=cmr6
\font\eighti=cmmi8 \font\sixi=cmmi6
\font\eightsy=cmsy8 \font\sixsy=cmsy6
\font\eightbf=cmbx8 \font\sixbf=cmbx6
\font\eightit=cmti8
\def\eightpoint{\def\rm{\fam0\eightrm}
  \textfont0=\eightrm \scriptfont0=\sixrm \scriptscriptfont0=\fiverm
  \textfont1=\eighti  \scriptfont1=\sixi  \scriptscriptfont1=\fivei
  \textfont2=\eightsy \scriptfont2=\sixsy \scriptscriptfont2=\fivesy
  \textfont3=\tenex   \scriptfont3=\tenex \scriptscriptfont3=\tenex
  \textfont\itfam=\eightit  \def\it{\fam\itfam\eightit}%
  \textfont\bffam=\eightbf  \scriptfont\bffam=\sixbf
  \scriptscriptfont\bffam=\fivebf  \def\bf{\fam\bffam\eightbf}%
  \normalbaselineskip=9pt
  \setbox\strutbox=\hbox{\vrule height7pt depth2pt width0pt}%
  \let\big=\eightbig  \normalbaselines\rm}
\catcode`@=11 %
\def\eightbig#1{{\hbox{$\textfont0=\ninerm\textfont2=\ninesy
  \left#1\vbox to6.5pt{}\right.\n@@space$}}}
\def\vfootnote#1{\insert\footins\bgroup\eightpoint
  \interlinepenalty=\interfootnotelinepenalty
  \splittopskip=\ht\strutbox %
  \splitmaxdepth=\dp\strutbox %
  \leftskip=0pt \rightskip=0pt \spaceskip=0pt \xspaceskip=0pt
  \textindent{#1}\footstrut\futurelet\next\fo@t}
\catcode`@=12 %
\def \de{\delta}

\def \si{\sigma}

\def \nab{\nabla}
\def \tnab{\tilde\nabla}
\def \pr{\partial}
\def \tr{{\rm tr }}

\def \hg{{\hat g}}
\def \hth{{\hat \theta}}

\def \l{\langle}
\def \r{\rangle}
\def \ep{\epsilon}
\def \vep{\varepsilon}
\def \half{{\textstyle {1 \over 2}}}
\def \thir{{\textstyle {1 \over 3}}}
\def \quar{{\textstyle {1 \over 4}}}
\def \ts{\textstyle}
\def \i{{\rm i}}
\def \d{{\rm d}}

\def \btau{{\bar \tau}}

\def \D{{\cal D}}

\def \L{{\cal L}}
\def \N{{\cal N}}

\def \U{{\cal U}}
\def \R{{\cal R}}
\def \Y{{\cal Y}}
\def \Z{{\cal Z}}

\font \bigbf=cmbx10 scaled \magstep1

\lref\hughtwo{J. Erdmenger and H. Osborn, {\it Conserved Currents and the 
Energy Momentum Tensor in Conformally Invariant Theories for General 
Dimensions}, Nucl. Phys. B483 (1997) 431, hep-th/9605009.}
\lref\hughone{H. Osborn and A. Petkou, {\it Implications of Conformal 
Invariance for Quantum Field Theories in $d>2$}
Ann. Phys. (N.Y.) {231} (1994) 311, hep-th/9307010.}

\lref\JackH{I. Jack and H. Osborn, {\it Analogs of the $c$-theorem for 
four dimensional renormalisable field theories}, Nucl. Phys. B343 (1990) 647.}
\lref\weyl{H. Osborn, {\it Weyl consistency conditions and a local 
renormalisation group equation for general renormalisable theories},
 Nucl. Phys. B363 (1991) 486.}
\lref\dewit{B.S. DeWitt, {\it Dynamical theory of groups and fields},
Blackie \& Son Ltd (Glasgow, 1965).}
\lref\Kap{V. Kaplunovsky and J Louis, {\it Field dependent gauge couplings in
locally supersymmetric quantum field theories},  Nucl. Phys. B422 (1994) 57, 
hep-th/9402005.}
\lref\Kraus{E Kraus, {\it Calculating the anomalous supersymmetry breaking in
Super-Yang-Mills theories with local coupling}, Phys. Rev. D65 (2002) 105003,
hep-th/0110323\semi
E Kraus, {\it An anomalous breaking of supersymmetry in supersymmetric gauge
theories with local coupling}, Nucl. Phys. B620 (2002) 55, hep-th/0107239\semi
E Kraus, C. Rupp and K Sibold, {\it Supercurrent and local coupling in the
Wess-Zumino model}, Eur. Phys. J. C24 (2002) 631, hep-th/0205013\semi
E Kraus, C. Rupp and K Sibold, {\it Supersymmetric Yang-Mills theories with
local coupling: the supersymmetric gauge}, hep-th/0212064.}
\lref\DZO{D.Z. Freedman and H. Osborn, {\it Constructing a $c$-function for 
SUSY gauge theories}, Phys. Lett. B 432 (1998) 353, hep-th/9804101.}
\lref\Tse{H. Liu and A.A. Tseytlin, {\it $D=4$ Super Yang Mills, $D=5$ gauged 
supergravity and $D=4$ conformal supergravity}, Nucl. Phys. B533 (1998) 88,
hep-th/9804083.}
\lref\Ven{A.E.M. van de Ven, {\it Two-loop quantum gravity}, Nucl. Phys.
B378 (1992) 309.}
\lref\HJ{I. Jack and H. Osborn, {\it Two-loop background field calculations
for arbitrary background fields}, Nucl. Phys. B207 (1982) 474.}

{\nopagenumbers
\rightline{DAMTP/03-11}
\rightline{hep-th/0302119}
\vskip 1.5truecm
\centerline {\bigbf Local Couplings and Sl(2,\Bbb R) Invariance for Gauge 
Theories at One Loop}
\vskip 1.5 true cm
\centerline {H. Osborn${}^\dagger$}

\vskip 12pt
\centerline {\ Department of Applied Mathematics and Theoretical Physics,}
\centerline {Silver Street, Cambridge, CB3 9EW, England}
\vskip 1.5 true cm

{\eightpoint
\parindent 1.5cm{

{\narrower\smallskip\parindent 0pt
The response of the one loop effective action for a gauge theory with local
couplings $g(x),\theta(x)$ under a local Weyl rescaling of the background metric
is calculated. Apart from terms which may be removed by local contributions to
the effective action the result is compatible with $Sl(2,R)$ symmetry acting on
$g,\theta$. Two loop effects are also discussed.



\narrower}}

\vfill
\line{${}^\dagger$ 
address for correspondence: Trinity College, Cambridge, CB2 1TQ, England\hfill}
\line{\hskip0.2cm email:
{{\tt ho@damtp.cam.ac.uk}\hfill}}
}

\eject}
\pageno=1

Requiring that a renormalisable quantum field theory is extended to be finite
for local couplings, so that the usual coupling constants are arbitrary
functions of position, ensures that functional derivatives of the effective 
action with respect to the local couplings directly define finite correlation 
functions for all composite operators that appear in the basic lagrangian. In 
supersymmetric theories local couplings, which may be given by supergravity
backgrounds, are directly relevant in constructing low energy effective theories and 
also for understanding the interplay of various anomalies \refs{\Kap,\Tse,\Kraus}. 
The use of local couplings also avoids the requirement in more conventional
treatments  of introducing additional local counterterms which are necessary for 
finiteness of composite operator correlation functions since they are
essentially included in the renormalised quantum action.
Further the usual renormalisation group equations can be extended to
equations corresponding to local Weyl rescalings of the metric. Imposing the
necessary integrability conditions has proved to
a convenient method for deriving non trivial relations amongst the $\beta$ 
and other functions that enter into the renormalisation of quantum field 
theories and the correlation functions of composite operators \weyl.
In two dimensions the necessary 
consistency relations involve identities which are equivalent, within
the allowed ambiguities, to the Zamolodchikov $c$-theorem and in four
dimensions similar results are obtained which are sufficient to show
irreversibility of renormalisation group flow in the perturbative regime
where the associated metric on the space of couplings can be calculated
and shown to be positive (unlike in two dimensions there is no presently
known connection to a two point function with manifest positivity properties).
The resulting equations then reflect some of the fundamental aspects of
quantum field theories.

Previously \JackH\ we calculated the various renormalisation quantities that
were necessary for a finite quantum field theory with local couplings
for general renormalisable theories in four dimensions at one and two loops.
In this note we extend these considerations to include the $\theta$ coupling
which is of course present in a general four dimensional gauge theory. 
Assuming a Euclidean metric $\gamma_{\mu\nu}(x)$, and arbitrary local couplings
$g(x),\theta(x)$ the gauge invariant  Euclidean action has the usual form
\eqn\action{
S = {1\over 4} \int \! \d^4x \sqrt \gamma \, \Big ( {1\over g^2} \, F^{\mu\nu}
{\cdot F_{\mu\nu}} - i \hth \,  F^{\mu\nu} {\cdot (*F)_{\mu\nu}}\Big )\, , \qquad
(*F)_{\mu\nu}= {1\over 2} {1\over\sqrt \gamma} \gamma_{\mu\tau}\gamma_{\nu\omega}
\ep^{\tau\omega\si\rho}F_{\si\rho}\, ,
}
with $F_{\mu\nu}$ the usual field strength and we define
\eqn\defth{
\hth = {\theta \over 8 \pi^2} \, .
}
For calculational convenience we choose a gauge fixing term of the form
\eqn\gfix{
S_{\rm g.f.} = {1\over 2} \int \! \d^4x \sqrt \gamma \, g^2 \, 
\nab^\mu \Big ( {1\over g^2} A_\mu \Big ) {\cdot \nab^\nu} \Big (
{1\over g^2} A_\nu \Big ) \, ,
}
and the associated ghost action is
\eqn\ghost{
S_{\rm gh} = \int \! \d^4x \sqrt \gamma \, {1\over g^2} \, \nab^\mu {\bar c}\,
{\cdot D_\mu} c \, , \qquad D_\mu c = \pr_\mu c + A_\mu \times c \, .
}
Adding $S$ and $S_{\rm g.f.}$ and expanding to quadratic order gives
\eqn\Squad{
S_{\rm quadratic} = {1\over 2} \int \! \d^4x \sqrt \gamma \, 
A^\mu {\cdot (\Delta_1 A)}_\mu  \, ,
}
where, using form notation\foot{Thus $\d$ is the exterior derivative with
$(\d A)_{\mu\nu} = \pr_\mu A_\nu - \pr_\nu A_\mu$ and 
for a scalar $\phi$ $(\d\phi)_\mu = \pr_\mu \phi$ while $\de$ is its adjoint with
$(\de F)_\mu = - {1\over \sqrt \gamma} \gamma_{\mu\omega}
\pr_\tau (\sqrt \gamma\gamma^{\tau\si} \gamma^{\omega\rho} F_{\si\rho})$,
$\de A = - {1\over \sqrt \gamma} \pr_\si (\sqrt \gamma
\gamma^{\si\rho} A_\rho)$. Also for a vector $u^\mu$ and a $n$-form 
$F_{\mu_1\dots \mu_n}$ we define $(\i_u F)_{\mu_1\dots \mu_{n-1}} =
u^\mu F_{\mu\mu_1\dots \mu_{n-1}}$.},
\eqn\Dvec{
\Delta_1 = \de {1\over g^2} \d + {1\over g^2} \d \, g^2 \de {1\over g^2}
+ \half i \, \i_{\pr \hth} * \d 
= \de {1\over g^2} \d + {1\over g^2} \d \, g^2 \de {1\over g^2} 
+ \half i \, \de \, \i_{\pr \hth} * \, .
}
If we introduce a modified connection
\eqn\conn{
\tnab_\si A_\mu = \nab_\si A_\mu + \half i g^2  \,
{1\over\sqrt \gamma} \gamma_{\si\tau}\gamma_{\mu\omega}
\ep^{\tau\omega\rho\nu} \pr_\rho \hth \, A_\nu \, ,
}
then $\Delta_1$ may be written in the form,
\eqn\dquad
{\eqalign{
\Delta_{1\mu}{}^{\!\nu} ={}& -\tnab^\si {1\over g^2} \tnab_\si \, 
\de_\mu{}^{\!\nu} + {1\over g^2} X_\mu{}^{\!\nu} \, , \cr
X_{\mu\nu} ={}& g^4 \nab_\mu {1\over g^2} \nab_\nu {1\over g^2}
- g^2 \nab_\mu \nab_\nu {1\over g^2} + \half g^4 \big (
\pr^\si \hth \pr_\si \hth\, \gamma_{\mu\nu}
- \pr_\mu \hth\, \pr_\nu \hth \big ) + R_{\mu\nu} \, . \cr}
}
The corresponding ghost operator
obtained from \ghost\ is
\eqn\Dghost{
\Delta_0 = \de {1\over g^2} \d =  - \nab^\si {1\over g^2} \nab_\si \, .
}
Using 
\eqn\rel{
g \nab^\si {1\over g^2} \nab_\si g = \nab^2 - X \, , \qquad
X = \half \, g^2 \nab^2 {1\over g^2} -  
\quar \, g^4 \nab^\si {1\over g^2} \nab_\si {1\over g^2} \, ,
}
we may easily see from \dquad\ and \Dghost,
\eqn\Op{
{\tilde \Delta}_1 = g \Delta_1 g = - \tnab{}^2 1 + Y_1 \, , \qquad
{\tilde \Delta}_0 = g \Delta_0 g = - \nab{}^2 + Y_0 \, ,
}
with $Y_{1\mu\nu} = X_{\mu\nu} + X \gamma_{\mu\nu}1, \ Y_0 = X1$.

The one loop effective action is then defined by
\eqn\Wloop{
W^{(1)} = - \half \ln \det {\tilde \Delta}_1 + \ln \det {\tilde \Delta}_0 \, .
}
For an operator $\Delta$ the functional determinant may be
defined in terms of the heat kernel,
\eqn\defdet{
- \ln \det \Delta = \zeta_\Delta\! {}'(0) \, , \qquad
 \zeta_\Delta (s) = {1\over \Gamma(s)} \int_0^\infty \!\!\! \d \tau \, 
\tau^{s-1}\,  \Tr \big ( e^{-\tau \Delta} \big ) \, .
}
We here consider the response to a local Weyl rescaling of the metric
$\de_\si \gamma_{\mu\nu} = 2\si \, \gamma_{\mu\nu}$ for this classically
conformally invariant theory. Using \Dvec\ and
\Dghost\ we have
\eqn\vDel{\eqalign{
\de_\si {\tilde \Delta}_1 = {}& - 2\si \, {\tilde \Delta}_1
+  2\si \, {1\over g}\d \, g^2 \de {1\over g} + {1\over g}
\d \, g^2 \de {1\over g} \, 2\si - {1\over g}\d \, 4 \si g^2 \de {1\over g}\, ,
\cr
\de_\si {\tilde \Delta}_0 ={}& - 4 \si \, g \, \de {1\over g^2} \d \, g +
g \, \de \, 2\si {1\over g^2} \d \, g \, , \cr}
}
so that, since $\de \Delta_1 = \Delta_0 g^2 \de{1\over g^2}, \ \Delta_1 \d
= {1\over g^2} \d g^2 \Delta_0$ and using cyclicity of the functional trace,
\eqn\varTr{
\de_\si \Big ( \Tr \big (  e^{-\tau {\tilde \Delta}_1 } \big ) -
2 \, \Tr \big (  e^{-\tau {\tilde \Delta}_0 } \big ) \Big ) = - 2\tau
{\d \over \d \tau} \Big ( \Tr \big ( \si e^{-\tau {\tilde \Delta}_1 } \big ) -
2 \,  \Tr \big ( \si  e^{-\tau {\tilde \Delta}_0 } \big ) \Big ) \, .
}
Hence with the definition \defdet\ we have
\eqn\vW{
\de_\si W^{(1)} = \Tr \big ( \si e^{-\tau {\tilde \Delta}_1 } \big ) 
\Big |_{\tau^0} - 2 \, \Tr \big ( \si e^{-\tau {\tilde \Delta}_0 } \big ) 
\Big |_{\tau^0} \, ,
}
where $|_{\tau^0}$ denotes the term ${\rm O}({\tau^0})$ in the heat kernel
expansion. For operators of the form $\Delta = - \nab^2 + Y $ we have
\eqn\siD{
\Tr \big ( \si e^{-\tau {\Delta} }  \big )\Big |_{\tau^0} = {1\over 16\pi^2} 
\int \!\d^4 x \sqrt \gamma \, \si \, \tr \big ( a_2^\Delta | \big ) \, ,
}
where the diagonal DeWitt coefficient is given by \dewit
\eqn\atwo{
a_2^\Delta | = {\ts{1\over 360}}(3F-G)1 + {\ts{1\over 12}}F^{\si\rho}F_{\si\rho}
+ \half \big ( Y - {\ts {1\over 6}} R \big )^2 -  {\ts {1\over 6}}
\nab^2 Y +  {\ts {1\over 30}} \nab^2 \! R \, 1\, .
}
with $F$ the square of the Weyl tensor, $G$ the Euler density, both
quadratic in the Riemann tensor\foot{$F=R^{\mu\nu\si\rho}R_{\mu\nu\si\rho}
- 2R^{\mu\nu}R_{\mu\nu} + {1\over 3}R^2, \ G =R^{\mu\nu\si\rho}R_{\mu\nu\si\rho}
-4 R^{\mu\nu}R_{\mu\nu} + R^2$.}, and $[\nab_\si, \nab_\rho]=F_{\si\rho}$.

Since according \Op\ both ${\tilde \Delta}_1$ and ${\tilde \Delta}_0$ are
of the required form we may readily obtain $a_2^{{\tilde \Delta}_1} |$
and $a_2^{{\tilde \Delta}_0} |$ using \atwo. For the former we may note
that
\eqnn\Yone
$$\eqalignno{
\tr_v \big ( (Y_1 - {\ts {1\over 6}} R )^2 \big ) = {}&
g^4 \nab^\mu \nab^\nu {1\over g^2}  \nab_\mu \nab_\nu {1\over g^2}
- 2g^6 \nab^\mu {1\over g^2}\nab^\nu {1\over g^2} \nab_\mu\nab_\nu {1\over g^2}\cr
&{}+ \half g^6 \nab^\mu {1\over g^2} \nab_\mu {1\over g^2} \nab^2 {1\over g^2}
+ {\ts {3\over 4}} g^8  \nab^\mu {1\over g^2} \nab_\mu {1\over g^2}
\nab^\nu {1\over g^2} \nab_\nu {1\over g^2} \cr
&{}+ \half g^6 \nab^2 {1\over g^2}\nab^\mu \hth \nab_\mu \hth  - g^6
\nab^\mu \nab^\nu {1\over g^2} \nab_\mu \hth \nab_\nu \hth + \quar g^8
\nab^\mu {1\over g^2} \nab_\mu {1\over g^2} \nab^\nu \hth \nab_\nu \hth \cr
&{} - g^8 \nab^\mu {1\over g^2}\nab^\nu {1\over g^2} \nab_\mu \hth \nab_\nu \hth 
+ {\ts {3\over 4}} g^8 \nab^\mu \hth \nab_\mu \hth \nab^\nu \hth \nab_\nu \hth \cr
&{}+ G^{\mu\nu} \Big ( 2 g^4 \nab_\mu {1\over g^2} \nab_\nu {1\over g^2} -
2 g^2 \nab_\mu\nab_\nu {1\over g^2} - g^4 \nab_\mu \hth \nab_\nu \hth \Big ) \cr
&{}+  R \Big ( \half  g^4 \nab^\mu {1\over g^2} \nab_\mu {1\over g^2} - 
\thir g^2 \nab^2 {1\over g^2} \Big ) + R^{\mu\nu}R_{\mu\nu} 
- {\ts {2\over 9}} R^2 \, , & \Yone
\cr}
$$
where $\tr_v$ denotes the trace over vector indices and $G_{\mu\nu} = R_{\mu\nu}
- \half \gamma_{\mu\nu}R$ is the Einstein tensor. In addition 
$F^{\si\rho}F_{\si\rho} \to - \R^{\mu\nu\si\rho}\R_{\mu\nu\si\rho}$
where $[\tnab_\si, \tnab_\rho] A_\mu =  \R_{\mu\nu\si\rho} A^\nu$ and
\eqnn\Rone
\hskip -1cm$$\eqalignno{
- \R^{\mu\nu\si\rho}\R_{\mu\nu\si\rho} = {}& 2 g^4 \nab^\mu \nab^\nu \hth
\nab_\mu \nab_\nu \hth + g^4 \nab^2 \hth \nab^2 \hth - 4 g^6
\nab^\mu {1\over g^2} \nab^\nu \hth \nab_\mu \nab_\nu \hth - 2  g^6
\nab^\mu {1\over g^2} \nab_\mu \hth \nab^2 \hth \cr
&{}+ 2 g^8 \nab^\mu {1\over g^2} \nab_\mu {1\over g^2} \nab^\nu \hth\nab_\nu \hth
+ g^8  \nab^\mu {1\over g^2} \nab^\nu {1\over g^2} \nab_\mu \nab_\nu \hth
- {\ts {3\over 4}} g^8  \nab^\mu \hth \nab_\mu \hth \nab^\nu \hth \nab_\nu\hth \cr
&{} + G^{\mu\nu} \, 2g^4 \nab_\mu \hth \nab_\nu\hth - 
R^{\mu\nu\si\rho}R_{\mu\nu\si\rho} \, .  & \Rone \cr}
$$

With these results it is straightforward to obtain from \vW, 
\eqn\siW{
16\pi^2 \de_\si W^{(1)} = \int \!\! \d^4 x \sqrt \gamma \, \si 
\Big ( \tr \big ( a_2^{{\tilde \Delta}_1}| \big ) - 2 \,
\tr \big ( a_2^{{\tilde \Delta}_0}| \big )\Big ) \, ,
}
where, for $n_V$ vector fields, we may write
\eqn\res{
\tr \big ( a_2^{{\tilde \Delta}_1}| \big ) - 2 \, 
\tr \big ( a_2^{{\tilde \Delta}_0}| \big ) = c F - a G - h \nab^2 R 
+ n_V \big ( \L - \nab_\mu \Z^\mu + \nab^2 \Y \big ) \, ,
}
which may be identified with $16\pi^2 \gamma^{\mu\nu}\l T_{\mu\nu}\r$ at one
loop. $\L$ can be reduced to the form
\eqn\resL{\eqalign{
\L ={}& \quar g^4 \bigg (\nab^2 {1\over g^2} \nab^2 {1\over g^2} +
\nab^2 \hth \nab^2 \hth - 2 G^{\mu\nu}\Big ( 
\nab_\mu {1\over g^2} \nab_\nu {1\over g^2} +\nab_\mu \hth\nab_\nu \hth \Big )\cr
&\qquad\quad {} - \thir R \Big (\nab^\mu {1\over g^2} \nab_\mu {1\over g^2}
+ \nab^\mu \hth \nab_\mu \hth \Big ) \bigg ) \cr
&{}- \half g^6 \Big ( \nab^\mu {1\over g^2} \nab_\mu {1\over g^2} \nab^2
{1\over g^2} + 2 \nab^\mu {1\over g^2} \nab_\mu \hth \nab^2 \hth - \nab^2
{1\over g^2} \nab^\mu \hth \nab_\mu \hth \Big ) \cr
&{}+ g^8 \Big ( {\ts {5\over 16}} \nab^\mu {1\over g^2} \nab_\mu {1\over g^2}
\nab^\nu {1\over g^2} \nab_\nu {1\over g^2} + {\ts {13\over 12}}
\nab^\mu {1\over g^2} \nab^\nu {1\over g^2} \nab_\mu \hth \nab_\nu \hth \cr
&\qquad\quad {} + {\ts {11\over 24}} \nab^\mu {1\over g^2} \nab_\mu {1\over g^2} 
\nab^\nu \hth \nab_\nu \hth + {\ts {5\over 16}} \nab^\mu \hth \nab_\mu \hth 
\nab^\nu \hth \nab_\nu \hth  \Big ) \, .  \cr}
}
In \res\ we should take $c = {1\over 10}n_V, \ a = {31\over 180}n_V, \
h = {1\over 10}n_V$ but
they may be left general since they are altered by one loop contributions
for scalar and spinor fields.
$\L$, $\Z^\mu$ and $\Y$, which depend on $\pr_\mu{1\over g^2}, \, \pr_\mu \hth$, 
are not so affected. For $\Z^\mu,\Y$ we also obtain
\eqn\resZY{\eqalign{
\Z^\mu = {}& G^{\mu\nu}g^2 \nab_\nu {1\over g^2} + \half \, g^4
\nab^\mu {1\over g^2} \nab^2 {1\over g^2} + {\ts{1\over 6}}\, g^4
\nab^\mu \hth \nab^2 \hth \cr
&{} - {\ts{1\over 2}}\, 
g^6 \nab^\mu {1\over g^2}\nab^\nu {1\over g^2}\nab_\nu {1\over g^2}
+ {\ts{1\over 12}}\, g^6 \nab^\mu {1\over g^2} \nab^\nu \hth \nab_\nu \hth
- \half\, g^6 \nab^\nu {1\over g^2} \nab^\mu \hth \nab_\nu \hth \, , \cr
\Y ={}& {\ts{1\over 6}}\, g^4 \nab^\mu {1\over g^2} \nab_\mu {1\over g^2}
- {\ts{1\over 6}}\, g^4 \nab^\mu \hth \nab_\mu \hth \, .  \cr}
}

We now demonstrate that $\L$ may be expressed in a form which exhibits manifest
$Sl(2,\Bbb R)$ invariance. We define as usual
\eqn\deft{
\tau = {\theta \over 2\pi} + i \, {4 \pi \over g^2} \, ,
}
where $\pmatrix{ a&b\cr c&d}\in Sl(2,\Bbb R)$ acts according to 
$\tau \to (a\tau +b)/(c\tau + d)$.
The quadratic terms in $\L$ correspond to the invariant line interval on the upper
half complex plane defined by
\eqn\met{
\d s^2 = {1\over ({\sl Im}\, \tau)^2}\, \d \tau \d \btau \, ,
}
which describes the constant negative curvature hyperboloid.
With this metric the Christoffel connections are just $\Gamma^\tau{}_{\! \tau\tau}
= i/{\sl Im}\, \tau, \ \Gamma^\btau{}_{\! \btau\btau} = - i
/{\sl Im}\, \tau$, and we may then define
\eqn\defDs{
D^2 \tau = \nab^2 \tau +{i \over {\sl Im}\, \tau}\nab^\mu \tau \nab_\mu\tau \, ,
\qquad D^2 \btau = \nab^2 \btau  - {i \over {\sl Im}\, \tau}
\nab^\mu \btau \nab_\mu\btau \, .
}
and consequently rewrite \resL\ in the form
\eqn\invL{\eqalign{
\L = {}& {1\over 4({\sl Im}\, \tau)^2} \big ( D^2 \tau D^2 \btau - 2 G^{\mu\nu}
\nab_\mu \tau \nab_\nu \btau - \thir R \, \nab^\mu \tau \nab_\mu \btau \big ) \cr
&{}+ {1\over 16({\sl Im}\, \tau)^4} \big ( \nab^\mu \tau \nab_\mu \btau \big )^2
+ {1\over 48 ({\sl Im}\, \tau)^4} \Big ( \nab^\mu \tau \nab^\nu \tau
\nab_\mu \btau \nab_\nu \btau - \big ( \nab^\mu \tau \nab_\mu \btau \big )^2 
\Big ) \, .  \cr}
}

The remaining terms given by \resZY\ cannot be expressed in an $Sl(2,\Bbb R)$
invariant form but they can be removed by taking 
$W^{(1)}\to W^{(1)}+W_{\rm loc}$ with,
\eqn\Wloc{\hskip -0.3cm \eqalign{
16\pi^2 W_{\rm loc} = {}& - {n_V\over 8} \int \!\! \d^4 x \sqrt \gamma \, 
\ln g^2 \, G\cr 
&{} -  n_V \! \int \!\! \d^4 x \sqrt \gamma \, \Big (
{\ts {1\over 8}} \, g^4 \nab^2 {1\over g^2} \nab^2 {1\over g^2} +
{\ts {1\over 24}} \, g^4 \nab^2 \hth \nab^2 \hth  - {\ts {1\over 6}}R \, \Y \cr
&\quad \ {} -  
\quar g^6  \nab^\mu {1\over g^2} \nab_\mu {1\over g^2} \nab^2 {1\over g^2} 
- \quar g^6   \nab^\mu {1\over g^2} \nab_\mu \hth \nab^2 \hth + 
{\ts {1\over 24}}g^6\, \nab^2 {1\over g^2} \nab^\mu \hth \nab_\mu \hth \Big )\, ,
\cr}
}
where we may easily calculate $\de_\si W_{\rm loc}$ using
$\de \nab^2 = - 2\si \nab^2 + 2\pr_\mu \si\nab^\mu, \
\de R = - 2\si R - 6 \nab^2 \si $ and $\de G = - 4 \si G  + 
8 G^{\mu\nu}\nab_\mu\pr_\nu \si$. Save for the $\ln g^2 \, G$ term, 
which cancels the term involving 
$G^{\mu\nu}\nab_\nu{1\over g^2}$ in \resZY, the terms present in $W_{\rm loc}$
reflect the usual arbitrariness up to local expressions in the effective action
arising from renormalisation scheme dependence\foot{In
\JackH\ a calculation for $\theta=0$ based on dimensional regularisation
gave different results for coefficients of the terms in $\Z^\mu, \Y$ except for
the one involving $G^{\mu\nu}$.}.

The possible presence of such additional local contributions may be seen 
by considering an alternative one loop effective action instead of \Wloop
\eqn\Whloop{
{\hat W}^{(1)} =  - \half \ln \det {\Delta}_1 + \ln \det {\Delta}_0 \, ,
}
where the operators $\Delta_1 ,\Delta_0$ are given by \Dvec\ and \Dghost,
without the rescaling in \Op\ which is equivalent in the original action
to taking $A_\mu = g a_\mu$ and correspondingly for the ghost fields. With
the zeta function definition \defdet\ we have for arbitrary $\lambda(x)$
\eqn\ldet{
\ln \det \lambda \Delta - \ln \det \Delta = 
\Tr \big ( \ln \lambda \ e^{-\tau {\Delta} }  \big )\Big |_{\tau^0} \, .
}
Hence we have
\eqn\WW
{ 16\pi^2\big ( {\hat W}^{(1)} - {W}^{(1)} \big ) = {1\over 2} 
\int \!\! \d^4x \sqrt \gamma \, \ln g^2 \, \big (  c F - a G - h \nab^2 R 
+ n_V  ( \L - \nab_\mu \Z^\mu + \nab^2 \Y ) \big ) \, .
}
The $\ln g^2 \, G$ term in \WW\ matches that in \Wloc\ if $a=\quar n_V$.
This is exactly the result for $\N=4$ supersymmetric gauge theories 
(when $c=a$). Under a Weyl rescaling 
$\de_\si \gamma_{\mu\nu} = 2\si \, \gamma_{\mu\nu}$
the remaining terms just lead to an expression for $\de_\si {\hat W}^{(1)}$
which is of the form \siW\ with the same result \resL\ for $\L$ although
a modified $\Z^\mu$. Exactly the same set of independent terms 
appear in the new $\Z^\mu$ as are present in the formula in \resZY.
As before these may then be cancelled by a simple local counterterm 
of the same form as \Wloc, with differing coefficients, but 
without the need now for a $\ln g^2 \, G$ term. To show this we need to
note that 
\eqn\siL{\eqalign{
& \de_\si \L = - 4\si \L + \nab_\mu ( \pr_\nu \si \, \U^{\mu\nu} ) \, , \cr
\U_{\mu\nu} = {}& g^4 \Big ( \nab_\mu {1\over g^2} \nab_\nu {1\over g^2} 
+ \nab_\mu \hth \nab_\nu \hth \Big ) - \half \gamma_{\mu\nu} \, g^4
\Big ( \nab^\rho {1\over g^2} \nab_\rho {1\over g^2} 
+ \nab^\rho \hth \nab_\rho \hth \Big ) \, . \cr}
}
The form for $\de_\si \L$ with $\U_{\mu\nu} = \U_{\nu\mu}$ follows from
the consistency relations in \weyl, this condition constrains the
quadratic and cubic terms in $\L$ although it is not fully determined.

It would of course be interesting to extend these considerations beyond one loop
and to see whether $Sl(2,\Bbb R)$ invariance is maintained.
In general we may write
\eqn\varW{ \eqalign{
16\pi^2 \D_\si W = {}& \int \!\! \d^4 x \sqrt \gamma \, \si \big (
c F - a G  - h \nab^2 R - {\ts {1\over 9}} b R^2
+ L - \nab_\mu Z^\mu + \nab^2 Y  \big ) \, , \cr
\D_\si = {}& \int \!\! \d^4x \, \si \Big ( - 2 \gamma^{\mu\nu} 
{\de \over \de \gamma^{\mu\nu}} + \beta^i {\de \over \de g^i } \Big ) \, , \cr}
}
for local couplings $g^i$, with corresponding $\beta$-functions $\beta^i$, and where
$L,Z^\mu,Y$ depend on their derivatives. Various consistency conditions were
derived in \weyl\ from $[\D_\si , \D_{\si'}]=0$. For a simple gauge coupling $g$,
with $\theta=0$, we may write
\eqn\gL{\eqalign {
L = n_V\bigg \{& {1\over g^2} \big ( \alpha \, (\nab^2 g)^2 - 2 \de \, G^{\mu\nu}
\pr_\mu g \pr_\nu g - {\ts {1\over 3}} \ep \, R \pr^\mu g \pr_\mu g \big ) \cr
\noalign{\vskip-4pt}
&{} - 2 \kappa \, {1\over g^3} \pr^\mu g \pr_\mu g \nab^2 g + 2 \lambda \, 
{1\over g^4} \pr^\mu g \pr_\mu g \pr^\nu g \pr_\nu g \bigg \} \, , \cr}
}
and to two loop order using dimensional regularisation, for $\hg^2 = g^2/16\pi^2$,
extending the results in \JackH\foot{The coefficient of $C$ in $\lambda$ is
corrected from \JackH.},
\eqn\resdim{\eqalign{
\alpha=\delta = {}& 1 + {\ts {1\over 3}} \big ( 51 C - 20 R_\psi - {\ts {7\over 2}}
R_\phi \big ) \hg^2 \, , \cr
\ep = {}& 1 + {\ts {1\over 3}} \big ( 29 C - 12 R_\psi - {\ts {5\over 2}}
R_\phi \big ) \hg^2 \, , \cr
\kappa = {}& 1 + {\ts {4\over 3}} \big ( 11 C - 4 R_\psi - {\ts {1\over 2}}
R_\phi \big ) \hg^2 \, , \cr
\lambda = {}& 1 + {\ts {1\over 18}} \big ( 323 C - 76 R_\psi - {\ts {25\over 2}}
R_\phi \big ) \hg^2 \, , \cr}
}
where it $t^\phi_a,t^\psi_a$ are the gauge group generators acting on scalar,
fermion fields, $\tr( t^\phi_a t^\phi_b ) = - \de_{ab} R_\phi , \
\tr( t^\psi_a t^\psi_b ) = - \de_{ab} R_\psi$. The results are scheme dependent.
For supersymmetric theories it is more natural to transform to a dimensional
reduction scheme by letting $1/\hg^2 \to 1/\hg^2 + {1\over 3}C$. For $\N=1$
supersymmetry we let $2R_\psi = C +R, \, R_\phi = 2R$ and also add ${2\over 3}R\hg^2$
to $\alpha,\delta,\ep$ since there are Yukawa couplings proportional to $g$ (for
more details see  \DZO). This gives
\eqn\resdimoone{\eqalign{
\alpha=\delta = {}& 1 + \big ( 13 C - 5 R\big ) \hg^2 \, , \qquad
\ep = 1 + \big ( 7 C - 3 R \big ) \hg^2 \, , \cr
\kappa = {}& 1 + 4\big ( 3 C -  R \big ) \hg^2 \, , \qquad \
\lambda = 1 + {\ts {1\over 2}} \big ( 31 C - 7 R \big ) \hg^2 \, . \cr}
}
For $\N=2$ theories $R\to C+2R$ and Yukawa couplings add a further $2C\hg^2$
to $\alpha,\delta,\ep$, giving now
\eqn\resdimtwo{\eqalign{
\alpha=\delta = {}& 1 + 10 \big ( C - R\big ) \hg^2 \, , \qquad
\ep = 1 + 6 \big ( C - R \big ) \hg^2 \, , \cr
\kappa = {}& 1 + 8\big ( C -  R \big ) \hg^2 \, , \qquad \
\lambda = 1 + \big ( 12 C - 7 R \big ) \hg^2 \, . \cr}
}
For $\N=4$, when there is a single adjoint hypermultiplet, $C=R$, as is necessary
for a zero $\beta$-function. In this case there are no corrections to
$\alpha,\delta,\ep,\kappa$, in accord with consistency relations, although
$\lambda$ remains non zero. Even for $\N=4$ there are thus additional perturbative
contributions beyond one loop. It is natural to suppose that for $\N=4$ $L$
in \gL\ extends with the inclusion of $\theta$ to a form which is invariant under 
$Sl(2,\Bbb Z)$, where $\alpha=\delta = \ep = \kappa =1$ and $\lambda$ becomes
an appropriate modular form.

\appendix{A}{Two loop Calculations with Local Couplings}

We here revisit old calculations for the divergences at two loops \JackH\
for pure gauge theories using just the local coupling $g$. Using integration 
by parts and Ward identities van der Ven \Ven\ showed that the two loop vacuum 
amplitude on flat space can be reduced to just
\eqn\start{
\eqalign{
W^{(2)} = {}& 
\int \!\! \int \! gg' \big ( - (D_\alpha G_{\beta\gamma}, G_{\beta\gamma}
{\overleftarrow D}{}'{}_{\! \delta}, G_{\alpha\delta} ) +
2 (D_\alpha G_{\beta\gamma},G_{\alpha\gamma}{\overleftarrow D}{\!}'{}_{\!\delta},
G_{\beta \delta} ) \big ) \cr
&{} - {1\over 4} \int \! g^2 \big ( 2 \tr (T_a G_{(\alpha\beta)}|
T_a G_{\alpha\beta}|) - \tr (T_a G_{\alpha\alpha}|T_a G_{\beta\beta}|)
+ \tr (T_a G_{\alpha\beta}|) \tr (T_a G_{\alpha\beta}|) \big ) \, , \cr}
}
where the ghost contribution is cancelled. In the first line of \start,
involving integrations over $x,x'$,
$G_{aa'\beta\gamma}(x,x')$ is the vector propagator,
$(X,Y,Z) = f_{abc}f_{a'b'c'}X_{aa'}Y_{bb'}Z_{cc'}$, $g=g(x), \, g'=g(x')$
are the local gauge couplings and
\eqn\defD{
D_\alpha = \pr_\alpha + v_\alpha(x) \, , \qquad
{\overleftarrow D}{\!}'{}_{\!\delta} = {\overleftarrow \pr}{\!}'{}_{\!\delta}
+ v_\delta(x') \, , \qquad v_\alpha = {1\over g} \pr_\alpha g \, .
}
In the second line of \start\ $(T_a)_{bc}=-f_{abc}$ and $G_{\alpha\beta}|$
denotes the coincident limit $x'=x$ and the trace is over group indices 
$a,b = 1, \dots , n_V$. The counterterms necessary to subtract sub-divergences 
are given by ($\tr\{T_a T_b\} = - C \de_{ab}$),
\eqnn\sub
$$\eqalignno{
W^{(2)}_{\rm c.t.} = - {C \over 16\pi^2\vep} \int \! g^2 \, \tr \big \{ &
{\ts {5\over 3}} D_\alpha G_{\beta\beta}{\overleftarrow D}{\!}'{}_{\!\alpha} |
- {\ts {14\over 3}} D_\alpha G_{\alpha\beta}
{\overleftarrow D}{\!}'{}_{\!\beta} | + 2 ( v_\beta D_\alpha G_{\alpha\beta} |
+ v_\alpha G_{\alpha\beta} {\overleftarrow D}{\!}'{}_{\!\beta} | ) \cr
& {} + 2 \pr{\cdot v} \, G_{\beta\beta}| - 4(\pr_\alpha v_\beta
+ v_\alpha v_\beta ) G_{\alpha\beta}| \big \} \, .  & \sub \cr}
$$
This form is in accord with that expected according to \JackH, combining the
ghost contribution here with the vector piece,\foot{
We note here the following misprints, in (5.7) the result for $Z^{(1)}_\beta$
should have a $-$ sign, in (5.1) $Y_{\rm gh}= - \nabla^{\si}v_\si + v^\si v_\si$.
Note also that the first two terms on the r.h.s. of (2.20) and the first three
on the r.h.s. of (2.21) should have the opposite sign.} 
if we note the identity
\eqn\ide{
D_\alpha G_{\alpha\beta} {\overleftarrow D}{\!}'{}_{\!\beta}| -
2 ( v_\beta D_\alpha G_{\alpha\beta} |
+ v_\alpha G_{\alpha\beta} {\overleftarrow D}{\!}'{}_{\!\beta} | ) 
+ 4 v_\alpha v_\beta  G_{\alpha\beta}|  = 0 \, .
}
Using an expansion of the propagators in terms of Seeley-DeWitt coefficients
$a_n(x,x')$ then in \HJ\ a simple algorithm was given for calculating the poles
in $\vep=4-d$,
\eqnn\div
$$\eqalignno{
W^{(2)}_{\rm div} = {}& {C \over (16\pi^2\vep)^2} \int \! g^2 \, \tr \big \{ 
({\ts {5\over 3}} + {\ts {\vep \over 36}} )  D_\alpha a_{1\beta\beta}
{\overleftarrow D}{\!}'{}_{\!\alpha} | - 
({\ts {14\over 3}} - {\ts {\vep \over 18}} )
D_\alpha a_{1\alpha\beta} {\overleftarrow D}{\!}'{}_{\!\beta} |  
+ (1 - {\ts {\vep \over 4}} ) a_{2\beta\beta}| \cr \noalign{\vskip-6pt}
&\qquad\qquad\qquad\ {} + 
( 2 - {\ts {\vep \over 2}}) (v_\beta D_\alpha a_{1\alpha\beta} |
+ v_\alpha a_{1\alpha\beta} {\overleftarrow D}{\!}'{}_{\!\beta} | ) 
+  ( 2 + {\ts {\vep \over 2}}) \pr{\cdot v} \, a_{1\beta\beta}| \cr
&\qquad\qquad\qquad\ {} - (4+\vep) \pr_\alpha v_\beta \, a_{1\alpha\beta}| - 
4 v_\alpha v_\beta \, a_{1\alpha\beta}|  +
{\ts {1 \over 4}}\vep \, a_{1\alpha\alpha}| \, a_{1\beta\beta}| \big  \}\cr
&{}+ {C \over (16\pi^2)^2 \vep} \int \! {\ts {1 \over 2}} \, \tr \big \{
\pr^2 g \pr^2 g \big \} \, . & \div \cr}
$$
Simplifying \div, using standard results for $a_1,a_2$, we get
\eqn\res{\eqalign{
W^{(2)}_{\rm div} = - {C n_V \over (16\pi^2\vep)^2} \int \! g^2 \, &
\Big \{ {\ts {22\over 3}}\big ( (\pr{\cdot v})^2  + 2 v^2 \pr{\cdot v} + 2v^2v^2 
\big ) \cr \noalign{\vskip-6pt}
&{} - \vep \big ( {\ts {17\over 2}} (\pr{\cdot v})^2 + {\ts {7\over 3}} 
v^2 \pr{\cdot v} + {\ts {106\over 9}} v^2v^2 \big ) \Big \} \, .
\cr}
}
Correspondingly at one loop
\eqn\resl{
W^{(1)}_{\rm div} = {n_V \over 16\pi^2\vep} \int \! \big ( (\pr{\cdot v})^2
+ v^2v^2 \big ) \, .
}
To two loop order this gives
\eqn\fin{\eqalign{
\Big ( \vep - {\hat \beta}(g){\pr \over \pr g} \Big ) W_{\rm div} =
{n_V \over 16\pi^2} \int \Big \{ & (1+17C {\hat g}^2 ) {1\over g^2}(\pr^2 g)^2
- 2( 1+ {\ts {44\over 3}} C {\hat g}^2 ) {1\over g^3} (\pr g)^2 \pr^2 g  \cr
\noalign{\vskip-6pt}
&{}+ 2 ( 1+ {\ts {323\over 18}} C {\hat g}^2 ) {1\over g^4} (\pr g)^2 (\pr g)^2 
\Big \} \, , \cr}
}
for ${\hat g}^2 = g^2/16\pi^2$.

\noindent{\bf Acknowledgements}

I would like to thank Johanna Erdmenger for discussions which prompted this
investigation, Massimo Bianchi for elucidating $Sl(2,\Bbb R)$ symmetry and Ian
Jack for recovering old calculations.
\listrefs
\bye